\documentclass{emulateapj}
\usepackage{amstext,amsmath,natbib}
\usepackage{epsfig}

\shorttitle{Ultracompact LMXBs}
\shortauthors{JUETT AND CHAKRABARTY}
\slugcomment{Accepted for publication in the Astrophysical Journal}

\begin{document}

\newcommand{\one}{4U~1850$-$087}
\newcommand{\two}{4U~0513$-$40}
\newcommand{\three}{4U~1822$-$000}
\newcommand{\four}{4U~1905$+$000}
\newcommand{\Ch}{{\em Chandra}}

\title{X-ray Spectroscopy of Candidate Ultracompact X-ray Binaries}

\author{Adrienne~M.~Juett\altaffilmark{1} and
Deepto~Chakrabarty}

\affil{\footnotesize Department of Physics and Center for Space
Research, Massachusetts Institute of Technology, Cambridge, MA
02139;\\ ajuett@virginia.edu, deepto@space.mit.edu}

\altaffiltext{1}{Current Address: Department of Astronomy, University
of Virginia, Charlottesville, VA 22903}

\begin{abstract}
We present high-resolution spectroscopy of the neutron star/low-mass
X-ray binaries (LMXBs) \one\/ and \two\/ as part of our continuing
study of known and candidate ultracompact binaries.  The LMXB \one\/
is one of four systems in which we had previously inferred an unusual
Ne/O ratio in the absorption along the line of sight, most likely from
material local to the binaries.  However, our recent {\em Chandra
X-ray Observatory} LETGS spectrum of \one\/ finds a Ne/O ratio by
number of 0.22$\pm$0.05, smaller than previously measured and
consistent with the expected interstellar value.  We propose that
variations in the Ne/O ratio due to source variability, as previously
observed in these sources, can explain the difference between the low-
and high-resolution spectral results for \one.  Our {\em XMM-Newton}
RGS observation of \two\/ also shows no unusual abundance ratios in
the absorption along the line of sight.  We also present spectral
results from a third candidate ultracompact binary, \three, whose
spectrum is well fit by an absorbed power-law $+$ blackbody model with
absorption consistent with the expected interstellar value.  Finally,
we present the non-detection of a fourth candidate ultracompact
binary, \four, with an upper limit on the source luminosity of
$<1\times10^{32}$~erg~s$^{-1}$.  Using archival data, we show that the
source has entered an extended quiescent state.
\end{abstract}


\section{Introduction}
Low-mass X-ray binaries (LMXBs) consist of a neutron star (NS) or
black hole (BH) accreting material from a $\lesssim$$1 \, M_{\odot}$
donor star.  One particularly interesting subset of the LMXB
population has orbital periods $\lesssim$80~min, the predicted minimum
period for binaries with hydrogen-rich main sequence donors.  Such
systems, commonly called ultracompact binaries, require
hydrogen-deficient or degenerate companions \citep{jar78,rj84,nrj86}.
Out of the 53 LMXBs with orbital period measurements, eight systems
have ultracompact periods.  The growing number of known and candidate
ultracompact binaries suggests that these systems make up a
significant fraction of the LMXB population.  We have undertaken X-ray
spectral observations of one known and two candidate ultracompact
binaries to better understand these unusual systems.

The LMXB 4U~1850$-$087 is an X-ray burster located in the globular
cluster NGC 6712.  It was detected by all of the major X-ray
satellites since {\em Uhuru} and shows a roughly factor of 10
variation in flux
\citep{fjc+78,hcl80,wmf+81,pt84,wmy+84,wnt+88,ktr92,cs97,jpc01}.
During the 1995 {\em ASCA} observation, the 0.5--10~keV X-ray flux was
$3.4 \times 10^{-10}$~erg~cm$^{-2}$~s$^{-1}$ \citep{jpc01}, which is
representative of the average source flux.  A low-amplitude 20.6-min
periodicity, attributed to the orbital period, was reported in the
{\em HST} data of the ultraviolet counterpart of 4U~1850$-$087
\citep{hcn+96}.  While this periodicity has yet to be confirmed, the
X-ray to optical luminosity ratio ($L_{\rm X}/L_{\rm opt}$) is
consistent with an ultracompact binary based on the relationship
between optical magnitude, X-ray luminosity, and orbital period
developed by \citet{vm94}.  Additionally, the X-ray luminosity of
4U~1850$-$087 is consistent with the expected mass-transfer rate for a
system with an orbital period of 21~min assuming a WD donor and
gravitational radiation driven mass transfer \citep{hcn+96}.

Spectral evidence also supports an ultracompact nature for
4U~1850$-$087.  \citet{spo+01} analyzed the {\em BeppoSAX} spectra of
the majority of the bright LMXBs located in globular clusters.  They
found that the best fit spectral parameters seemed to fall into two
groups, with one group consisting of 4U~1820$-$30, 4U~1850$-$087, and
4U~0513$-$40, i.e., the known or candidate ultracompact binaries.
Additionally, we found that the {\em ASCA} spectrum of 4U~1850$-$087
revealed an unusual feature at 0.7~keV also seen in three other
candidate ultracompact binaries \citep{jpc01}.  We attributed this
feature to an unusual Ne/O ratio in the interstellar absorption edges,
presumably influenced by material local to the binary.  However, a
recent observation of \one\/ using {\em XMM-Newton} found no evidence
for an unusual Ne/O ratio \citep{spo04}.

The LMXB 4U~0513$-$40 is an X-ray burster in the globular cluster NGC
1851.  The binary was first discovered by the MIT {\em OSO}-7
satellite and has been studied by all of the major X-ray instruments
that followed
\citep{cml75,jcc+77,fjc+78,mlp+81,hg83,sl92,cpc95,s99,spo+01}.  The
flux from \two\/ shows a factor of ten variability from observation to
observation.  During the 2000 {\em BeppoSAX} observation, \two\/ had a
0.1--100~keV X-ray luminosity of $5 \times 10^{36}$~erg~s$^{-1}$
\citep{spo+01}, an average flux for the source.  \citet{dma00}
identified \two\/ as a candidate ultracompact binary based on its
$L_{\rm X}/L_{\rm opt}$ ratio.  In addition, its broadband optical,
ultraviolet, and X-ray spectrum resembles that of other ultracompact
LMXBs \citep{dma00,spo+01}.

We identified the Galactic field LMXB \three\/ as an ultracompact
binary candidate based on its $L_{\rm X}/L_{\rm opt}$ ratio.  The
source was discovered by {\em Uhuru} and has been observed by all of
the major X-ray missions
\citep{gmg+72,mwl+79,wmf+81,wmy+84,sl92,gpr+95,cs97}.  The source
\three\/ has been roughly constant in flux over the course of X-ray
astronomy, with a 0.5--20~keV unabsorbed flux of $1\times
10^{-9}$~erg~cm$^{-2}$~s$^{-1}$ \citep{cs97}.  It has a faint optical
counterpart (V$=$22) similar to other candidate ultracompact binaries
\citep{ci85}.

In the published literature, \four\/ was found to be persistently
bright, with a 2--10~keV flux of
$3\times10^{-10}$~erg~cm$^{-2}$~s$^{-1}$, and showed Type-I X-ray
bursts \citep{llh+76,spt+76,fjc+78,rjb+80,cic85,ci90,gpr+95,cs97}.
These observations were performed in the 1970s and 1980s.  The source
was observed by more recent X-ray satellites, but the results of these
observations have never been published.  From the analysis of a radius
expansion burst, \citet{ci90} estimated the distance to \four\/ at
8$\pm$1~kpc.  The optical counterpart found by \citet{ci85} is a
V$=$20.5, ultraviolet-excess star.

In this paper, we present the results from X-ray spectral observations
of \one, \two, and \three.  The non-detection of \four\/ is also
presented.  The details of the observations and data reduction
techniques are given in \S\ref{sec:2}, and the analysis is presented
in \S\ref{sec:3}.  In \S\ref{sec:4}, we discuss the implications of
these results and summarize the case for neon-rich donors in
ultracompact binaries.

\section{Observations and Data Reduction}\label{sec:2}
\subsection{{\em Chandra} Observation of 4U~1850$-$087}
We observed \one\/ with the {\em Chandra X-ray Observatory} on 2002
August 16 for 50 ks using the Low Energy Transmission Grating
Spectrometer (LETGS) and the Advanced CCD Imaging Spectrometer
\citep[ACIS;][]{wbc+02}.  The LETGS spectra are imaged by ACIS, an
array of six CCD detectors.  The LETGS/ACIS combination provides both
an undispersed (zeroth order) image and dispersed spectra from the
grating with a first order wavelength range of 1.4--63~\AA\/
(0.2--8.9~keV) and a spectral resolution of $\Delta\lambda=$
0.05~\AA\/ ($R$$=$$\lambda/\Delta\lambda$$=$28--1260).  The various
orders overlap and are sorted using the intrinsic energy resolution of
the ACIS CCDs.  The observation of \one\/ used a Y-offset of
$+$1\farcm5 in order to place the O-$K$ absorption edge on the
back-side illuminated S3 CCD, which has suffered less degradation than
the front-side illuminated CCDs.  The detected zeroth and first order
count rates for \one\/ were 1.2 and 3.0 cts~s$^{-1}$, respectively.

The ``level 1'' event file was processed using the CIAO v3.0 data
analysis package\footnote{http://asc.harvard.edu/ciao/}.  The standard
CIAO spectral reduction procedure was performed.  We filtered the
event file retaining those events tagged as afterglow events by the
{\tt acis\_detect\_afterglow} tool.  Since order-sorting of grating
spectra provides efficient rejection of background events, the
afterglow detection tool is not necessary to detect cosmic ray
afterglow events.  No features were found that might be attributable
to afterglow events.  For bright sources, pileup can be a problem for
CCD detectors \citep[see e.g.,][]{d03}.  The zeroth order {\em
Chandra} spectrum of \one\/ was heavily affected by pileup and was not
used in this analysis.  The first order spectrum suffered from minimal
pileup in the 6--8~\AA\/ range (1.5--2.1~keV) which has the effect of
making the instrumental iridium edge feature more pronounced.
Detector response files (ARFs and RMFs) for the plus and minus first
order spectra were created using the standard CIAO tools.

Calibration studies have determined that a layer of contaminant is
building on the optical blocking filter of ACIS, absorbing incoming
X-rays.  The contaminant gradually reduces the detector quantum
efficiency with time and affects all wavelengths, particularly those
above 6~\AA.  The {\em Chandra} instrument teams have studied the
spectral profile of the contaminant absorption using LETGS spectra and
have modeled its structure and time-dependence \citep{mtg+03}.  An
estimated correction for the contaminant is now available through the
{\em Chandra} website\footnote{See
http://asc.harvard.edu/ciao/threads/aciscontam/}.  The biggest effect
is an overall reduction in effective area due to the large optical
depth of carbon.  In addition, small edges from oxygen and fluorine
are also detected.  The ARFs used in this analysis have been corrected
for the effect of the contaminant.

The plus and minus first order spectra and response files were summed
to provide a single first order spectrum and response.  Background
spectra were extracted from the standard LETG background regions.
After background subtraction, the data were binned to a resolution of
0.075~\AA\/ to ensure good statistics.  Spectral analysis of the {\em
Chandra} observation of \one\/ was performed using ISIS \citep{hd00}.

\subsection{{\em Chandra} Observation of 4U~1822$-$000}
{\em Chandra} observed \three\/ on 2003 August 3 for 2~ks using the
High Energy Transmission Grating Spectrometer (HETGS) and ACIS
\citep{wbc+02}. The HETGS carries two transmission gratings: the
Medium Energy Gratings (MEGs) with a range of 2.5--31~\AA\/
(0.4--5.0~keV) and the High Energy Gratings (HEGs) with a range of
1.2--15~\AA\/ (0.8--10.0~keV).  The HETGS spectra are imaged by ACIS,
which provides both an undispersed (zeroth order) image and dispersed
spectra from the gratings.  The various orders overlap and are sorted
using the intrinsic energy resolution of the ACIS CCDs.  The
first-order MEG (HEG) spectrum has a spectral resolution of
$\Delta\lambda=$ 0.023~\AA\/ or $R$$=$110--1350 (0.012~\AA\/ or
$R$$=$100--1250).  The detected zeroth and first order count rates for
\three\/ were 0.7 and 11.6 cts~s$^{-1}$, respectively.

The level 1 event file for the \three\/ observation was processed
using CIAO v3.0.  The zeroth order {\em Chandra} spectrum of \three\/
was heavily affected by pileup (pileup fraction $>$80\%) and was not
used in this analysis.  We checked the dispersed spectra of \three\/
and found that pileup was minimal ($<$10\%).  The standard CIAO tools
were used to create ARFs and RMFs for the MEG and HEG $+1$ and $-1$
order spectra.  The ARFs include the contamination correction
mentioned previously.  The ARFs were combined when the plus and minus
first order spectra were added for the MEG and HEG separately.  We
created background spectra for the MEG and HEG using the standard
background regions.  The data were binned to ensure at least 50 counts
per bin.  The {\em Chandra} spectral analysis of \three\/ was
performed using XSPEC v11.3 \citep{a96}.

%\subsection{{\em Chandra} Observation of \four}
%We observed \four\/ on 2003 March 4 for 5~ks with {\em Chandra} in the
%HETG/ACIS configuration.  The level 1 event file for \four\/ was
%processed using CIAO v3.0.  No source was found at the position of
%\four.  The {\tt celldetect} source detection algorithm was used to
%search for any sources in the field of view.  No sources were found.

\subsection{{\em XMM-Newton} Observation of 4U~0513$-$40}
The {\em XMM-Newton Observatory} observed \two\/ on 2003 April 1 for
24~ks.  {\em XMM-Newton} carries three different instruments, the
European Photon Imaging Cameras \citep[EPIC;][]{sbd+01,taa+01}, the
Reflection Grating Spectrometers \citep[RGS;][]{hbk+01}, and the
Optical Monitor \citep[OM;][]{mbm+01}.  The OM data was not used in
this analysis.  The EPIC instruments consist of three CCD cameras, MOS
1, MOS 2, and pn, which provide imaging, spectral, and timing data.
The pn camera was run in small window mode and the two MOS cameras
were run in full window mode.  The EPIC instruments provide good
spectral resolution ($\Delta E$$=$50--200~eV FWHM) over the 0.3--12.0
keV range.  There are two RGS onboard {\em XMM-Newton} which provide
high-resolution spectra ($R$$=$100--500 FWHM) over the 5--38 \AA\/
(0.33--2.5~keV) range.  The grating spectra are imaged onto CCD
cameras similar to the EPIC-MOS cameras which allows for order sorting
of the high-resolution spectra.  The detected pn count rate for \two\/
was 20.8 cts~s$^{-1}$, while the first order combined RGS count rate
was 2.3 cts~s$^{-1}$.

The {\em XMM-Newton} data were reduced using the Science Analysis
System (SAS) version 5.4.1.  Standard filters were applied to all {\em
XMM-Newton} data.  The EPIC-pn data were reduced using {\tt epchain}.
The pn spectrum was extracted from a region 45\arcsec\/ in radius
around the source.  Response files were generated for the pn spectrum
using {\tt rmfgen} and {\tt arfgen}.  We reduced the EPIC-MOS data
using {\tt emchain}.  The MOS data were found to have considerable
pileup and were therefore not used in this analysis.  The RGS data
were reduced using {\tt rgsproc}, which produced standard first order
source and background spectra and response files for both RGS.  We
grouped the pn spectrum to oversample the energy resolution of the CCD
by no more than a factor of three.  The RGS spectra were grouped to
ensure that each bin had at least 20 counts.  All {\em XMM-Newton}
spectral analysis of \two\/ was performed using XSPEC \citep{a96}.

\begin{figure}
\centerline{\epsfig{file=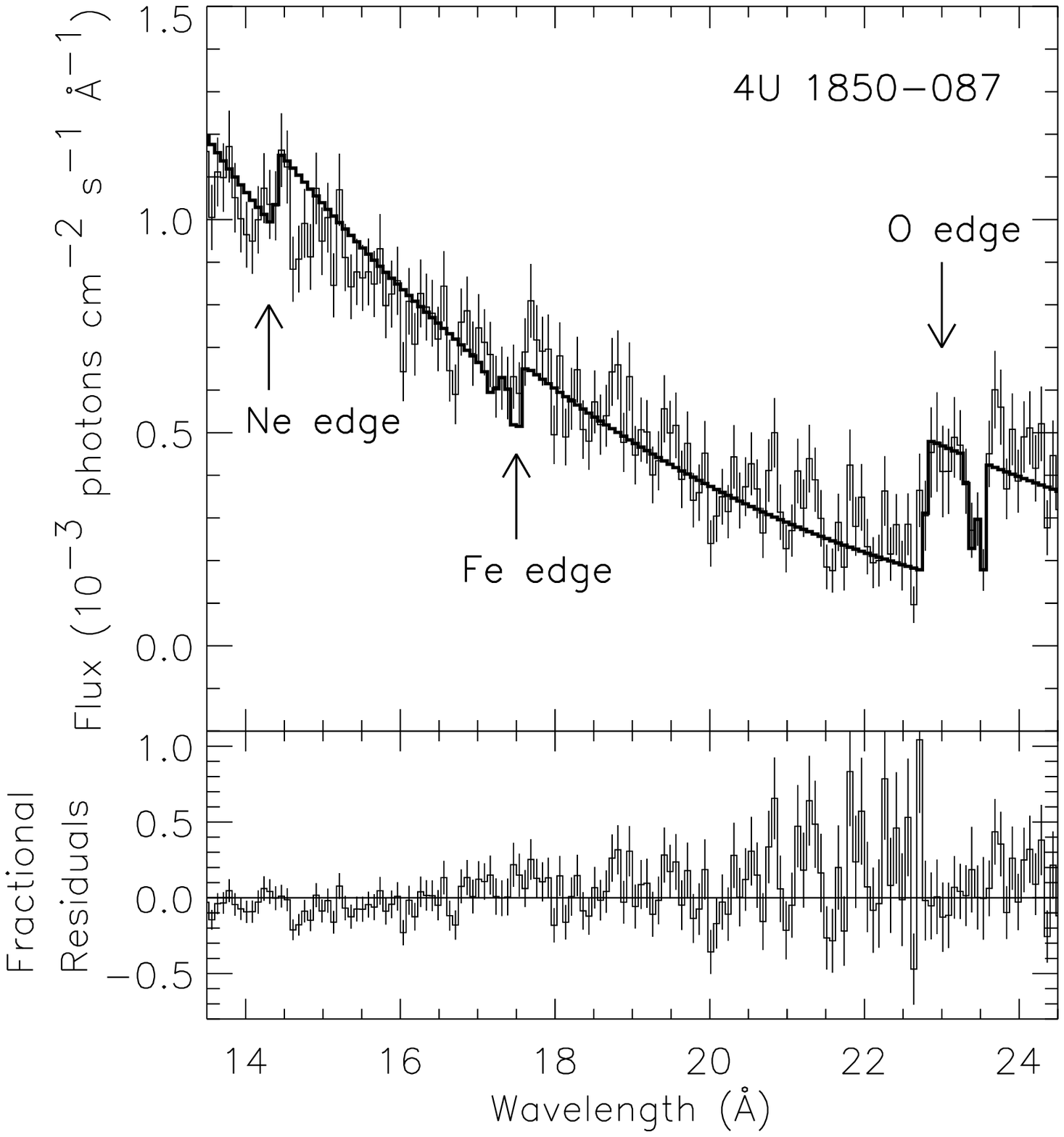, width=0.95\linewidth}} 
\caption{(Top panel) Combined LETG first order spectrum of \one\/ with
best-fit model including the oxygen, iron, and neon edge models.
(Bottom panel) Fractional residuals ([data-model]/model) of the
spectral fits shown above.  The absorption lines at 23.5 and
23.3~\AA\/ are the expected interstellar absorption lines from
\ion{O}{1} and \ion{O}{2}.  There are no other features detected which
are consistent with narrow emission or absorption lines.  The arrows
mark the neon, iron, and oxygen absorption edges at 14.3, 17.5, and
23~\AA\/ respectively.}
\label{fig:1850}
\end{figure} 

\section{Analysis and Results}\label{sec:3}
\subsection{{\em Chandra} Observation of 4U~1850$-$087}
We previously showed that \one\/ was one of four sources to have a
feature at 0.7~keV in its {\em ASCA} spectrum which could be explained
by an unusual Ne/O ratio in the absorbing material along the line of
sight \citep{jpc01}.  High-resolution spectroscopy of the other three
sources has confirmed the unusual Ne/O ratios \citep{pbv+01,jc03}.  To
investigate the absorption along the line of sight to \one, we fit the
first order LETGS spectrum over the 1.5--25.5~\AA\/ wavelength range
with a model that separately fit for the column densities of the
oxygen, iron, and neon edges.  The 5--10~\AA\/ wavelength range was
excluded from the fits in order to avoid the piled up region of the
spectrum.  All errors are 90\%-confidence unless otherwise noted.

\begin{deluxetable*}{lccccccc}
\tabletypesize{\footnotesize} 
\tablewidth{0pt} 
\setlength{\tabcolsep}{0.1in}
\tablecaption{Best-Fit Continuum Spectral Parameters\tablenotemark{a}}
\tablehead{\colhead{Source} & \colhead{$\Gamma$} & 
  \colhead{$A_{1}$\tablenotemark{b}} & \colhead{$kT_{\rm bb}$ (keV)} & 
  \colhead{$R^2_{\rm km}/D^2_{\rm 10 \, kpc}$} & 
  \colhead{Flux\tablenotemark{c}} & \colhead{$f_{\rm pl}$\tablenotemark{d}} & 
  \colhead{$\chi^2_{\nu}/ \nu$} } 
\startdata
\one & 1.48$\pm$0.05 & 1.32$\pm$0.09 & 0.479$\pm$0.009 & 113$\pm$13 & 1.45 & 
    0.66 & 1.3/242 \\
\two & 1.963$\pm$0.017 & 2.04$\pm$0.06 & 0.75$\pm$0.08 & 2.0$\pm$0.7 & 0.97 & 
    0.96 & 1.12/1612 \\
\three & 1.3$\pm$0.3 & 8$\pm$2 & 0.86$^{+0.11}_{-0.07}$ & 46$^{+26}_{-17}$ & 
    8.4 & 0.73 & 1.47/452 
\enddata
\tablenotetext{a}{All errors quoted at the 90\%-confidence level.}
\tablenotetext{b}{Power-law normalization in units of $10^{-2}$
photons cm$^{-2}$ s$^{-1}$ keV$^{-1}$.}  
\tablenotetext{c}{The absorbed 0.5--10~keV flux in units of
$10^{-10}$~erg~cm$^{-2}$~s$^{-1}$.}  
\tablenotetext{d}{Fraction of the absorbed flux in the power-law
component.}
\label{tab:1}
\end{deluxetable*}

The continuum model consisted of a power law $+$ blackbody and
absorption by the {\tt tbvarabs} model with the abundances for oxygen,
iron, and neon set to zero and all others equal to the interstellar
abundances of \citet{wam00}.  The equivalent hydrogen column density
$N_{\rm H}$ in {\tt tbvarabs} was set to the expected interstellar
value of $2\times 10^{21}$~cm$^{-2}$ derived from the reddening
$E_{\rm B-V}$ to the cluster \citep{kdi+03} and using the relationship
between $E_{\rm B-V}$ and $N_{\rm H}$ of \citet{ps95}.  The $K$-shell
edges of oxygen and neon were fit using a standard {\tt edge} model,
while the Fe-$L$ edge was fit using a custom multiplicative model
based on the cross-sections of \citet{kk00}.  Two Gaussian absorption
lines were included at the oxygen edge to represent the $1s$-$2p$
absorption lines from \ion{O}{1} and \ion{O}{2}.  For oxygen and neon,
the atomic absorption cross-sections were taken from the theoretical
calculations of \citet{gm00} and \citet{g00}, respectively.  The
spectrum of \one\/ and best-fit spectral model including the
absorption edges is shown in Figure~\ref{fig:1850}.  The best-fit
power-law and blackbody parameters are given in Table~\ref{tab:1}.
The absorbed 0.5--10~keV flux for \one\/ was
$1.45\times10^{-10}$~erg~cm$^{-2}$~s$^{-1}$ with the power-law
component contributing 66\% of the flux.  At a distance of 8.2~kpc
\citep{kdi+03}, the absorption corrected 0.5--10~keV luminosity of
\one\/ was $1.35 \times 10^{36}$~erg~s$^{-1}$, assuming isotropic
emission.

The best-fit absorption edge and line values are given in
Table~\ref{tab:2}.  The best-fit edge positions are consistent with
the expected positions of the edges.  We find a Ne/O ratio along the
line of sight to \one\/ of 0.22$\pm$0.05, consistent with the expected
interstellar value of 0.18 using the ISM abundances of \citet{wam00},
as well as that found in the {\em XMM-Newton} spectrum \citep{spo04}.
The oxygen column density implies a hydrogen column density of $N_{\rm
H} = (4.2\pm0.4)\times10^{21}$~cm$^{-2}$ assuming standard ISM
abundances \citep{wam00} in agreement with the predicted $N_{\rm H}$
from the neon and iron edges, although a factor of two larger than the
expected $N_{\rm H}$ of $2\times 10^{21}$~cm$^{-2}$ from the reddening
to the cluster \citep{kdi+03}.  Other than the interstellar oxygen
absorption lines, no other emission or absorption lines were apparent
in the spectrum of \one.  We performed a careful search of the {\em
Chandra} spectral residuals to place limits on the presence of any
spectral features in the 12--25~\AA\/ wavelength range.  Gaussian
models with fixed FWHM~$=2000$~km~s$^{-1}$ were fit at each point.
(This width is similar to that seen in emission lines from other
LMXBs, e.g., 4U~1626$-$67, \citealt{scm+01}, and EXO~0748$-$67,
\citealt{ckb+01}).  From this, we can place a $3\sigma$ upper limit on
the EW of any line feature, either emission or absorption
\citep[see][for a cautionary note]{pvc+02}.  The EW limit increases
with wavelength, varying from 0.05~\AA\/ at 12~\AA\/ to 0.25~\AA\/ at
24~\AA.

The LETGS first order dispersed spectrum of \one\/ has an average
count rate of $2.80\pm0.17$~cts~s$^{-1}$.  We examined the total count
rate, as well as the count rates in two different energy ranges, to
check for changes in the spectral state.  We found no evidence for any
change of state during the {\em Chandra\/} observation.  To look for
periodic modulations of the X-ray flux, we created lightcurves from
event files after barycentering and randomizing the event arrival
times.  (Randomizing of the event arrival times consists of adding a
random quantity uniformly distributed between zero and the readout
time of 0.74~s in order to avoid aliasing caused by the readout time.)
We searched for modulations of the X-ray flux with frequencies between
1$\times10^{-5}$ and 5$\times10^{-2}$ Hz.  We found no evidence for
periodic modulation with a 90\%-confidence upper limit of 1.3\% for
the fractional rms amplitude.

To determine the position of \one, we first applied the aspect offset
correction available at the {\em Chandra} website\footnote{See
http://asc.harvard.edu/cal/ASPECT/celmon/index.html}.  Using the CIAO
tool {\tt celldetect}, the zeroth order source position for \one\/ was
determined: R.A.=$18^{\rm h} 53^{\rm m} 04\fs86$ and Dec=$-8^{\circ}
42\arcmin 20\farcs4$, equinox J2000.0 (90\% confidence error radius of
0\farcs6).  This position is 0\farcs6 from the optical counterpart
position given by \citet{amd+93}.

\subsection{{\em XMM-Newton} Observation of 4U~0513$-$40}
To determine the appropriate continuum model for \two, we fit the pn
and RGS spectra simultaneously with either an absorbed power-law or an
absorbed power-law $+$ blackbody model.  The pn spectrum was fit over
the energy range 1.5--12~keV, while the two RGS spectra were fit over
the range 0.34--2.0~keV.  Residuals on the order of 10\% were seen in
the pn at energies between 0.5--1.5~keV.  While this is consistent
with estimates of the effective area calibration for the pn, given the
high count rate of \two, these residuals resulted in inflated
$\chi^{2}_{\nu}$ values so we chose to exclude this region of the pn
spectrum from our fits.  A constant was also included to allow for
normalization differences between the pn and RGS.  The absorbed
power-law $+$ blackbody model was a significantly better fit
(significance of 98\% as calculated by an $F$-test) to the data and we
adopt this as the appropriate continuum model.  The best-fit power-law
and blackbody parameters are given in Table~\ref{tab:1}.  The absorbed
0.5--10~keV flux for \two\/ was
$9.66\times10^{-11}$~erg~cm$^{-2}$~s$^{-1}$ with the power-law
component contributing 96\% of the flux.  At a distance of 12.1~kpc
\citep{kdi+03}, this gives a luminosity of $1.8 \times
10^{36}$~erg~s$^{-1}$.

\begin{figure} 
\centerline{\epsfig{file=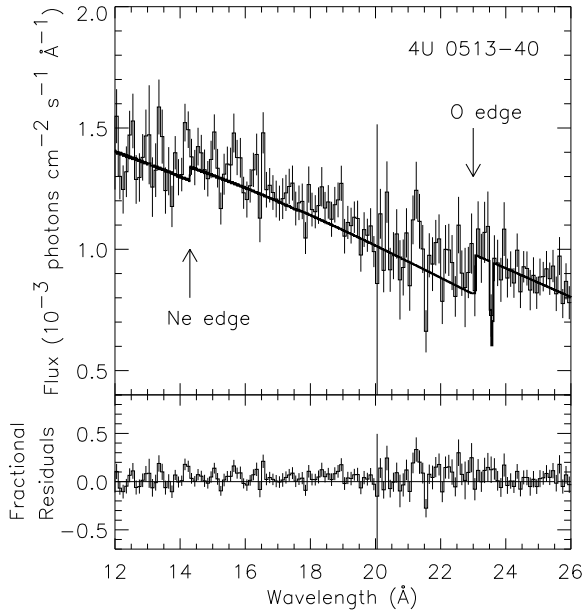,width=0.95\linewidth}}
\caption{(Top panel) Combined RGS first order spectrum of \two\/ with
best-fit model including the oxygen, iron, and neon edge models.
(Bottom panel) Fractional residuals ([data-model]/model) of the
spectral fits shown above.  The absorption line at 23.5~\AA\/ is the
expected interstellar absorption line from \ion{O}{1}.  There are no
other features detected which are consistent with narrow emission or
absorption lines.  The arrows mark the neon and oxygen absorption
edges at 14.3 and 23~\AA\/ respectively. }
\label{fig:0513}
\end{figure}

To determine the column density along the line of sight to \two, we
fit the RGS spectra with separate models for the absorption edges from
oxygen, iron, and neon as was done for \one.  The continuum model was
fixed to that found in the pn and RGS combined fit.  A Gaussian
absorption line was included to model the $1s$-$2p$ absorption line
from \ion{O}{1}.  The spectrum of \two\/ and best-fit spectral model
are shown in Figure~\ref{fig:0513}.  The best-fit absorption line and
edge parameters are given in Table~\ref{tab:2}.  The absorption to
\two\/ is very low ($N_{\rm H} \sim 10^{20}$~cm$^{-2}$) making a
determination of both the wavelength and depth of the iron and neon
edges difficult.  We therefore fixed the iron and neon edge
wavelengths to the expected values of 17.5 and 14.3~\AA, respectively.
Given the strength of the oxygen edge complex, its position was
allowed to vary and the best-fit edge wavelength is comparable to what
is found in other LMXBs \citep{jsc04}.  The oxygen edge measurement
implies an equivalent hydrogen column of $N_{\rm H} =
(7.6\pm1.2)\times 10^{20}$~cm$^{-2}$ assuming standard ISM abundances
\citep{wam00}.  The neon and iron edges are consistent with this
value.  The X-ray derived $N_{\rm H}$ value is greater than the value
of $N_{\rm H} = 1.1 \times 10^{20}$~cm$^{-2}$ derived from the
reddening to the cluster \citep{kdi+03}.  No emission or absorption
lines were apparent in the spectrum of \two.  We performed a search of
the {\em XMM-Newton} RGS spectral residuals over the 6--36~\AA\/
wavelength range.  The $3\sigma$ upper limit on the EW of any spectral
feature is roughly constant at 0.1~\AA\/ over the range 6--30~\AA\/
and then increases to 0.4~\AA\/ at 35~\AA\/ for Gaussian models with
fixed FWHM~$=2000$~km~s$^{-1}$ \citep[see][for a cautionary
note]{pvc+02}.

The \two\/ observation had average count rates of 20.8$\pm$1.2 and
7.1$\pm$0.6 cts s$^{-1}$ for the pn and combined RGS, respectively.
Using the pn data, we checked for changes in the spectral state by
examining the total count rate, as well as the count rates in two
different energy ranges.  There was no evidence for a change in the
spectral state of \two\/ during the {\em XMM-Newton} observation.  Due
to the low count rate of this observation ($\ll 1$ count per readout),
the low frequency end of the power spectrum is affected by
instrumental noise \citep[see][and references therein]{kkb+02}.  To
remove the noise, we calculated the average power at each frequency
using an 11 bin moving average and then divided by this value.  We
verified that the resulting powers were consistent with an exponential
distribution, and then searched for modulation of the X-ray flux with
frequencies between $3\times10^{-4}$ and 1 Hz.  No modulation was
detected with a 90\%-confidence upper limit on the fractional rms
amplitude of 0.7\%.

\subsection{{\em Chandra} Observation of 4U~1822$-$000}
We fit the {\em Chandra} HEG and MEG first-order spectra jointly with
both an absorbed power-law and an absorbed power-law $+$ blackbody
model over the wavelength ranges 1.8--11.3~\AA\/ for the HEG and
2.1--14.3~\AA\/ for the MEG.  The data were binned to have at least 50
counts per bin which given the low number of counts from the source
resulted in significantly reduced spectral resolution. This combined
with the high $N_{\rm H}$ along the line of sight resulted in few
counts being detected shortward of 0.9~keV.  Therefore, it was not
necessary to use the complex absorption model presented earlier.
Instead, we used the interstellar absorption model of \citet[{\tt
tbabs};][]{wam00} with the abundances in XSPEC set to {\tt wilm} and
the cross-section set to {\tt vern}.  The power-law $+$ blackbody
model produced a significantly better chi-squared value than the
power-law alone (significance of 98.4\% as calculated by an $F$-test),
so we take a power-law $+$ blackbody model as the appropriate
continuum model for the {\em Chandra} spectra.  The best-fit
equivalent hydrogen column density was $N_{\rm H} = (0.97\pm0.18)
\times 10^{22}$~cm$^{-2}$ which compares well with the value derived
from infrared dust maps \citep[$N_{\rm H} \approx 0.8 \times
10^{22}$~cm$^{-2}$;][]{sfd98}.  The best-fit power-law and blackbody
parameters are given in Table~\ref{tab:1}.  The absorbed 0.5--10~keV
flux for 4U~1822$-$000 was $8.4\times10^{-10}$~erg~cm$^{-2}$~s$^{-1}$
with the power-law component contributing 73\% of the flux.

The combined MEG $+$ HEG first order dispersed spectrum of
4U~1822$-$000 has an average count rate of $11.3\pm0.5$~cts~s$^{-1}$.
We examined the total count rate, as well as the count rates in two
different energy ranges, to check for changes in the spectral state.
We found no evidence for any change of state during the {\em
Chandra\/} observation.  To look for periodic modulations of the X-ray
flux, we created lightcurves from event files after barycentering and
randomizing the event arrival times.  To increase statistics, we
included events from the first through tenth grating orders.  We
searched for modulations of the X-ray flux with frequencies between
2$\times10^{-4}$ and 5$\times10^{-2}$ Hz.  We found no evidence for
periodic modulation with a 90\%-confidence upper limit of 2.6\% for
the fractional rms amplitude.

\begin{deluxetable*}{lccc}
%\tabletypesize{\footnotesize} 
\tablewidth{0pt} 
\setlength{\tabcolsep}{0.1in}
\tablecaption{Best-Fit Absorption Feature Parameters\tablenotemark{a}}
\tablehead{\colhead{Feature} & \colhead{$\lambda$ (\AA)} & 
  \colhead{$N_{\rm X}$ ($10^{18}$ cm$^{-2}$)} & 
  \colhead{EW (m\AA)} }
\startdata
\multicolumn{4}{c}{\one} \\ \tableline
O-$K$ edge & $22.79^{+0.05}_{-0.11}$ & $2.1\pm0.2$ & \nodata \\
\ion{O}{1} $1s$-$2p$ line & $23.511\pm0.004$ & \nodata & $70\pm30$ \\
\ion{O}{2} $1s$-$2p$ line & $23.363^{+0.04}_{-0.014}$ & \nodata & $40\pm40$ \\
Fe-$L$ edge & $17.52\pm0.07$ & $0.08\pm0.03$ & \nodata \\
Ne-$K$ edge & $14.40\pm0.08$ & $0.46\pm0.10$ & \nodata \\ \tableline
\multicolumn{4}{c}{\two} \\ \tableline
O-$K$ edge & $23.05^{+0.13}_{-0.69}$ & $0.37\pm0.06$ & \nodata \\
\ion{O}{1} $1s$-$2p$ line & $23.4^{+3.5}_{-0.9}$ & \nodata & $31\pm26$ \\
Fe-$L$ edge & 17.5 (fixed) & $<0.014$ & \nodata \\
Ne-$K$ edge & 14.3 (fixed) & $0.10\pm0.08$ & \nodata
\enddata
\tablenotetext{a}{All errors quoted at the 90\%-confidence level.}
\label{tab:2}
\end{deluxetable*}

To determine the position of 4U~1822$-$000, we first applied the
aspect offset correction.  Using the CIAO tool {\tt tgdetect}, the
zeroth order source position for 4U~1822$-$000 was determined:
R.A.=$18^{\rm h} 25^{\rm m} 22\fs02$ and Dec=$-00^{\circ} 00\arcmin
43\farcs0$, equinox J2000.0 (90\% confidence error radius of
0\farcs6).  This position is 0\farcs9 away from the optical
counterpart position \citep{gsp+91}.

\subsection{{\em Chandra} Observation of \four}
We observed \four\/ on 2003 March 4 for 5~ks with {\em Chandra} in the
HETG/ACIS configuration.  We did not detect a source during this
observation.  From the non-detection, we place an upper limit on the
flux of $1\times10^{-14}$~erg~cm$^{-2}$~s$^{-1}$ assuming a 0.3~keV
blackbody and $N_{\rm H}= 1.9\times 10^{21}$~cm$^{-2}$.  From a radius
expansion burst, \citet{ci90} derived a distance to \four\/ of
8$\pm$1~kpc.  At this distance, the upper limit on the source
luminosity is $1\times10^{32}$~erg~s$^{-1}$, making \four\/ one of the
faintest known quiescent NS systems.

In an effort to determine when the source went into quiescence, we
retrieved the archival data from {\em ROSAT} and {\em ASCA}
observations of \four.  {\em ROSAT} observations of Aql X-1 would have
included \four\/ in the field of view.  We retrieved the two PSPCB
observations of Aql X-1 performed on 1992 October 15 and 1993 March
24.  No source at the position of \four\/ was found with a upper limit
on the flux of $6\times10^{-13}$~erg~cm$^{-2}$~s$^{-1}$.  Similarly,
the {\em ASCA} observation of \four\/ performed on 1995 April 09 found
no source at the position of \four.  From this work, we conclude that
\four\/ went into its quiescent state in the late 1980s or early 1990s
after an extended outburst phase.

\section{Discussion}\label{sec:4}
In this paper, we have shown that the {\em Chandra}/LETGS spectrum of
\one, the {\em XMM-Newton}/RGS spectrum of \two, and the {\em
Chandra}/HETGS spectrum of \three\/ are all well fit by an absorbed
power-law $+$ blackbody model.  No unusual abundance ratios are
implied by the detected absorption features, and no emission or
absorption features were found other than the expected absorption
lines due to interstellar oxygen.  For \one, this is significantly
different from what is implied by the earlier {\em ASCA} observation,
which had established \one\/ as being part of a class of four NS/LMXBs
all having a similar feature at 0.7~keV in their low-resolution
spectra \citep{jpc01}.  This feature had originally been attributed to
unresolved line emission from iron and oxygen \citep[see
e.g.,][]{cws94,wka97}.  However, a high-resolution observation of the
brightest of these sources, 4U~0614$+$091, with {\em Chandra}, failed
to detect any emission lines, finding instead an unusually high Ne/O
number ratio in the absorption along the line of sight \citep{pbv+01}.
We had subsequently pointed out that the {\em ASCA} spectra of all
four sources are well fit without a 0.7~keV emission line using a
model that includes photoelectric absorption due to excess neon along
the lines of sight and presumably local to the sources \citep{jpc01}.

Of the other four sources that we previously proposed as neon-rich
systems based on their low-resolution {\em ASCA} spectra, three
(4U~0614$+$091, 2S~0918$-$549, and 4U~1543$-$624) have also shown
unusual Ne/O abundance ratios in their high-resolution X-ray spectra
\citep{pbv+01,jc03}.  However, a comparison of the {\em Chandra}, {\em
XMM-Newton}, and {\em ASCA} results for 2S~0918$-$549 and
4U~1543$-$624 also revealed variability in the measured Ne/O ratio,
which seemed to be associated with changes in the source continuum
properties \citep{jc03}.  We previously suggested that changes in the
spectral properties of the sources could lead to ionization effects
that would produce the variable Ne/O ratio found for the neutral
edges.  Ionization would affect oxygen more than neon, inflating the
measured neutral Ne/O ratio.  Unfortunately, this means that the
measured Ne/O ratio would not reflect the donor composition.

\begin{figure}
\centerline{\epsfig{file=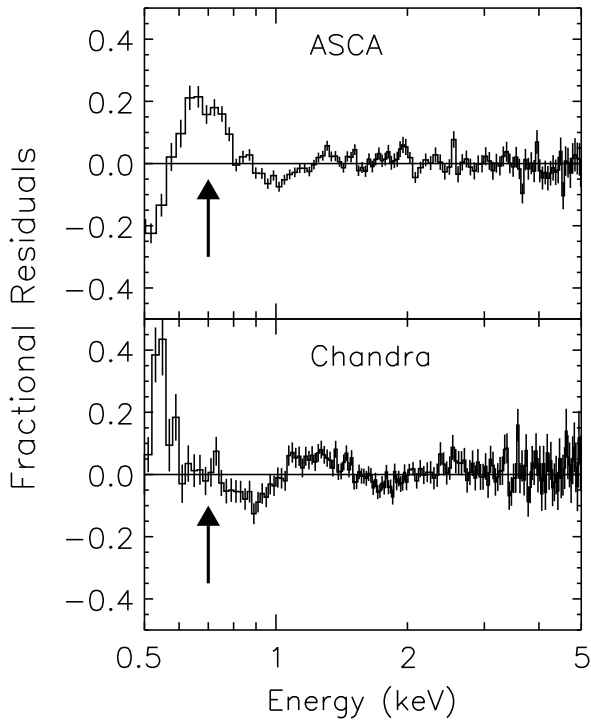, width=0.98\linewidth}}
\caption{Comparison of the {\em ASCA} (top panel) and {\em Chandra}
(bottom panel) spectrum for \one\/ plotted at the same energy
resolution.  Shown are the fractional residuals ([data $-$
model]/model) of an absorbed power law $+$ blackbody model.  The {\em
ASCA} spectrum shows a residual feature at 0.7~keV (marked by an
arrow) which is absent in the {\em Chandra} data.  This feature was
previously attributed to an unusual Ne/O abundance ratio in the
absorption.  The lack of the 0.7~keV feature in the {\em Chandra}
spectrum supports our suggestion that the source spectrum has changed
between the two observations.  The residual in the {\em Chandra}
spectrum at 0.55~keV is due to misalignment of the oxygen edge between
the data and model.  }
\label{fig:comp}
\end{figure}

Here we propose that for \one\/ as well, ionization effects due to
continuum spectral variations are also responsible for the differences
between the {\em ASCA} and {\em Chandra} results.  When the {\em
Chandra} spectrum is rebinned to a resolution comparable to that of
{\em ASCA}, the {\em Chandra} data do not show the 0.7~keV feature
seen previously with {\em ASCA} (see Figure~\ref{fig:comp}).  This
clearly indicates that the source spectrum changed between 1995 and
2002.  Given our interpretation of the feature as being due to an
unusual Ne/O ratio in the absorbing material, the lack of the 0.7~keV
feature in the {\em Chandra} spectrum is consistent with the measured
Ne/O ratio of 0.22$\pm$0.05, which is close to the interstellar value.
Had the feature still been present in the binned {\em Chandra}
spectrum, then our interpretation would have been incorrect.  The
variation of the Ne/O ratio in the absorbing material is accompanied
by differences in the continuum spectral properties of \one.  The {\em
Chandra} observation of \one\/ was taken during a lower luminosity
state than the {\em ASCA} observation which showed the unusual Ne/O
ratio.  This is comparable to what was observed in the other proposed
neon-rich sources 2S~0918$-$549 and 4U~1543$-$624.  A more detailed
analysis of the {\em XMM-Newton} data may shed more light on this
source.  Given that the {\em XMM-Newton} results show a Ne/O ratio
similar to that found in the {\em Chandra} data \citep{spo04}, we
would expect that the continuum properties should also be similar.

While X-ray spectra may be inconclusive in determining the donor
composition of these systems, new results from optical spectroscopy
have confirmed their unusual nature.  Optical spectra of three
candidate ultracompact binaries with suggested neon-rich donors,
4U~0614$+$091, 4U~1543$-$624, and 2S~0918$-$549 \citep{jpc01}, showed
no prominent hydrogen or helium emission lines \citep{njm+04,w04}.  In
the brightest of these, 4U~0614$+$091, \citet{njm+04} found emission
lines which they attributed to carbon and oxygen.  The other spectra
were not as high quality, but hints of similar spectral features were
seen.  The optical spectra seem to confirm our suggestion that the
donors in these systems are C-O WDs \citep{jpc01,njm+04}.  The notion
of a class of ultracompact LMXBs with neon-rich C-O WD donors is
motivated by the definite detection of one such system, 4U~1626$-$67
\citep{scm+01,haw+02}.  However, not all ultracompact binaries are
expected to have C-O donors.  The burst properties of the 11-min
binary 4U~1820$-$30 support a helium WD donor for that system, while
the properties of the 50-min X-ray dipper 4U~1916$-$05 point to a
hydrogen-deficient but not yet degenerate donor.  As more observations
allow for the donor compositions of the ultracompact binaries to be
determined, the results will provide a unique test of binary evolution
theory. Multiwavelength observations may be the best way of
determining the donor properties in ultracompact binary systems.

\acknowledgements{Partial support for this work was provided by NASA
through Chandra award numbers GO2-3065X and GO3-4055X issued by the
Chandra X-Ray Observatory Center, which is operated by the Smithsonian
Astrophysical Observatory for NASA under contract NAS8-03060.  This
work was also supported in part by NASA under grants NAG5-13179 and
NAG5-9184 as well as contract NAS8-01129.  This research made use of
data obtained through the High Energy Astrophysics Science Archive
Research Center (HEASARC), operated by the NASA/Goddard Space Flight
Center.}

%\bibliography{/home/ajuett/papers/thesis/mybib}
%\bibliographystyle{apj}

\end{document}